\documentclass[12pt]{article}

\usepackage{graphicx}
\usepackage{graphics}
\usepackage{color}
\usepackage{ragged2e}
\usepackage{lipsum}
\usepackage{amsmath}
\usepackage{cite}
\usepackage{url}

\def\beq{\begin{equation}}
\def\eeq{\end{equation}}
\def\bea{\begin{eqnarray}}
\def\eea{\end{eqnarray}}
\def\nn{\nonumber}
\def\roughly#1{\mathrel{\raise.3ex\hbox
{$#1$\kern-.75em\lower1ex\hbox{$\sim$}}}}

\def\gsim{\roughly>}

\def\sla#1{\raise.15ex\hbox{$/$}\kern-.57em #1}

\def \cB{{\cal B}}
\def \SM{{\rm SM}}
\def\bsmumu{b \to s \mu^+ \mu^-}

\def\bctaunu{b \to c \tau^- {\bar\nu}}
\def\s{\sqrt{2}}
\def \SM{{\rm SM}}
\def \NP{{\rm NP}}

\begin{document}

\begin{flushright}
UdeM-GPP-TH-19-275 \\
\end{flushright}

\begin{center}
\bigskip
{\Large \bf \boldmath Anomalies in $B$ Decays: A Sign of New
  Physics?\footnote{Talk given at the XI$^{\rm th}$ International
    Symposium on Quantum Theory and Symmetries, July 1-5, Centre de
    recherches math\'ematiques, Montr\'eal, Canada}} \\
\bigskip
\bigskip
{\large
David London\footnote{london@lps.umontreal.ca}
}
\end{center}

\begin{center}
{\it Physique des Particules, Universit\'e de Montr\'eal,}\\
{\it C.P. 6128, succ. centre-ville, Montr\'eal, QC, Canada H3C 3J7}
\end{center}

\begin{center}
\bigskip (\today)
\vskip0.5cm {\Large Abstract\\} \vskip3truemm
\parbox[t]{\textwidth}{At the present time, there are a number of
  measurements of $B$-decay observables that disagree with the
  predictions of the standard model. These discrepancies have been
  seen in processes governed by two types of decay: (i) $\bsmumu$ and
  (ii) $\bctaunu$. In this talk, I review the experimental results, as
  well as the proposed new-physics explanations. We may be seeing the
  first signs of physics beyond the standard model.}

\end{center}


\thispagestyle{empty}
\newpage
\setcounter{page}{1}
\baselineskip=14pt

\section{Introduction}

The development of the standard model (SM) in particle physics is one
of the great triumphs in all of physics.  The SM has made a great many
predictions, almost all of which have been verified, including the
existence of the Higgs boson. There is no question that the SM is
correct. 

However, there are many reasons to believe it is not complete, such as
the large number of arbitrary parameters, the hierarchy problem, the
matter-antimatter asymmetry in the universe, dark matter, etc. In
order to address these issues, there must be physics beyond the SM.
We don't know what the new physics (NP) is, nor where it is, so we
have to search for it in all possible ways:
\begin{itemize}

\item[$\bullet$] Direct searches: in high-energy experiments, one task is to look
  for the production of new particles. Unfortunately, to date, such
  searches have revealed nothing. No SUSY, no direct dark matter
  detection, no new particles.

\item[$\bullet$] Indirect searches: here the idea is to look for virtual effects
  of new particles. This method has been more promising.

\end{itemize}

\section{\boldmath $B$-Decay Anomalies}

\subsection{$\bsmumu$}

\begin{figure}[b]
\begin{center}
\includegraphics[width=0.4\hsize]{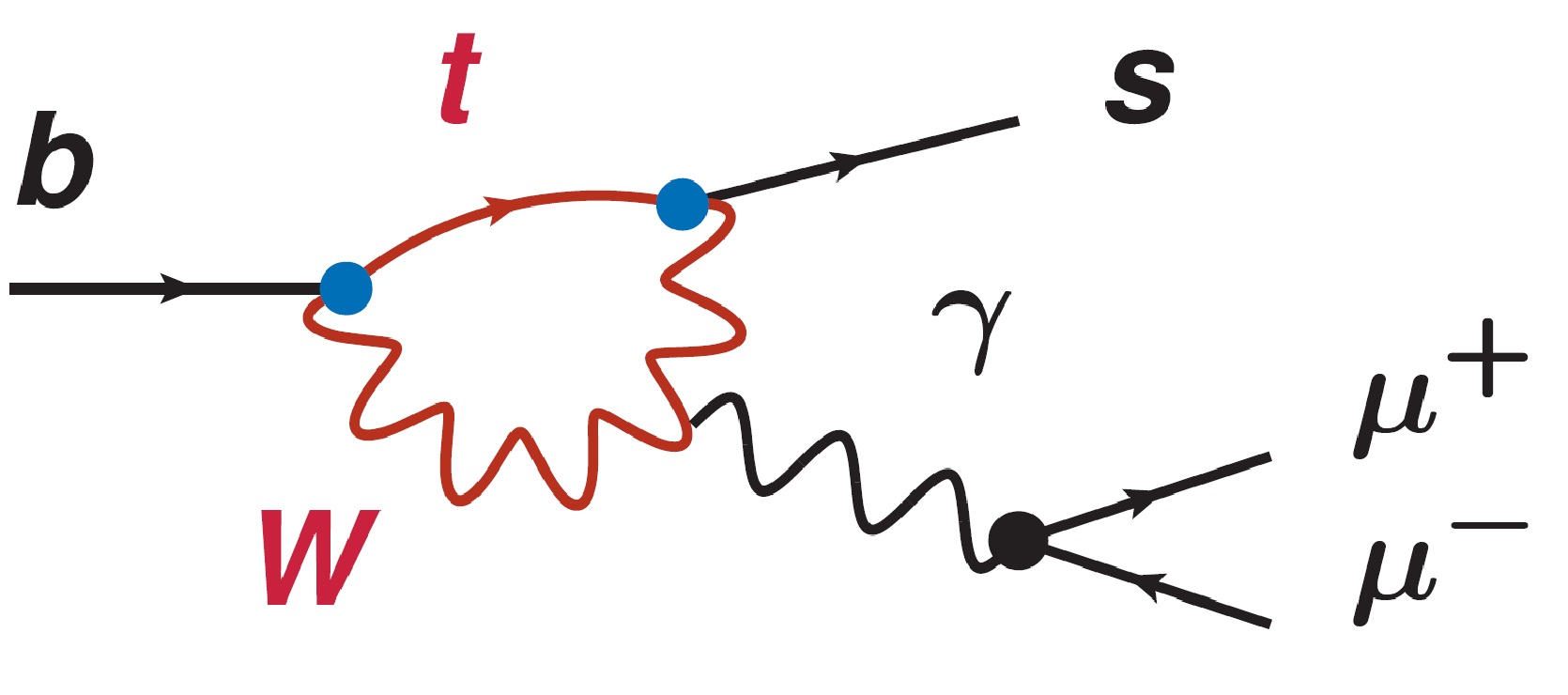}
\caption{SM diagram for $\bsmumu$.}
\label{bsmumufig}
\end{center}
\end{figure}

$b \to s$ transitions, which have $\Delta Q_{em} = 0$, are
flavour-changing neutral-current (FCNC) processes. In the SM, these
can arise only at loop level. One such FCNC decay is $\bsmumu$, the
diagram for which is shown in Fig.~\ref{bsmumufig}. The SM amplitude
is suppressed by loop factors and small elements in the
Cabibbo-Kobayashi-Maskawa (CKM) quark mixing matrix:
\beq
A \sim {\frac{1}{16\pi^2}} \frac{g^4}{M_W^2} \frac{m_t^2}{M_W^2} {V_{tb} V_{ts}^*} ~.
\eeq

Processes whose rates are small in the SM are excellent places to
search for NP. Indeed, there are a number of measurements of
observables involving $\bsmumu$ that disagree with the predictions of
the SM:
\begin{itemize}

\item[$\bullet$] The measured branching ratios of $B \to K^* \mu^+ \mu^-$
  \cite{Aaij:2014pli} and $B_s \to \phi \mu^+ \mu^-$
  \cite{Aaij:2015esa} have been found to be smaller than the
  predictions of the SM. Here there are significant theoretical
  uncertainties, related to the poorly-known values of the hadronic
  form factors \cite{Straub:2015ica,Horgan:2015vla,Gubernari:2018wyi}.

\item[$\bullet$] Deviations from the SM expectations have been found
  in measurements of the angular distribution of $B \to K^* \mu^+
  \mu^-$
  \cite{Aaij:2015oid,BK*mumuATLAS,BK*mumuCMS,Khachatryan:2015isa},
  particularly in the angular observable $P'_5$ \cite{P'5}. Here, the
  form-factor uncertainties are smaller than for the branching ratios
  \cite{Khodjamirian:2010vf,Bobeth:2017vxj}, but they are still
  important.

\item[$\bullet$] LHCb has measured 
\beq
R_K \equiv \frac{\cB(\bar{B} \to K \mu^+ \mu^-)}{\cB(\bar{B} \to K e^+ e^-)} ~.
\eeq
Using the Run 1 data (2014) \cite{RKexpt}, for $1 \le q^2 \le 6~{\rm
  GeV}^2$, where $q^2$ is the dilepton invariant mass-squared, it was
found that
\beq
R_{K,{\rm Run~1}}^{\rm old} = 0.745^{+0.090}_{-0.074}~{\rm (stat)} \pm 0.036~{\rm (syst)} ~.
\eeq
The SM prediction is $R_K^\SM = 1 \pm 0.01$ \cite{IsidoriRK}.  This
measurement disagrees with the SM at the level of $2.6\sigma$,
suggesting a violation of lepton universality.

At the Rencontres de Moriond, 2019, LHCb presented new $R_K$ results
\cite{LHCbRKnew}: (i) the Run 1 data were reanalyzed using a new
reconstruction selection method, and (ii) the Run 2 data were
analyzed. The results are
\bea
R_{K,{\rm Run~1}}^{\rm new} &=& 0.717^{+0.083}_{-0.071}~{\rm (stat)} ^{+0.017}_{-0.016}~{\rm (syst)} ~, \nn\\
R_{K,{\rm Run~2}} &=& 0.928^{+0.089}_{-0.076}~{\rm (stat)} \pm ^{+0.020}_{-0.017}~{\rm (syst)} ~. 
\eea
Combining the Run 1 and Run 2 results gives
\beq
R_K = 0.846^{+0.060}_{-0.054}~{\rm (stat)} ^{+0.016}_{-0.014}~{\rm (syst)} ~.
\label{RKmeas}
\eeq
The central value is closer to the SM prediction, but, due to the smaller errors,
the discrepancy with the SM is still $\sim 2.5\sigma$.

LHCb has also measured 
\beq
R_{K^*} \equiv \frac
{\cB(\bar{B} \to K^* \mu^+ \mu^-)}{\cB(\bar{B} \to K^* e^+ e^-)} ~,
\eeq
finding \cite{RK*expt}
\beq
R_{K^*} = 
\begin{cases}
0.660^{+0.110}_{-0.070}~{\rm (stat)} \pm 0.024~{\rm (syst)} ~,~~ 0.045 \le q^2 \le 1.1 ~{\rm GeV}^2 ~, \\
0.685^{+0.113}_{-0.069}~{\rm (stat)} \pm 0.047~{\rm (syst)} ~,~~ 1.1 \le q^2 \le 6.0 ~{\rm GeV}^2 ~. 
\end{cases} \nn
\eeq
Compared to SM predictions, these correspond to discrepancies of
$2.4\sigma$ and $2.5\sigma$.

At the Rencontres de Moriond, 2019, Belle announced its measurement of
$R_{K^*}$ \cite{BelleRK*new}:
\beq
R_{K^*} =
\begin{cases}
0.52^{+0.36}_{-0.26} \pm 0.05 ~,~~ 0.045 \le q^2 \le 1.1 ~{\rm GeV}^2 ~, \\
0.96^{+0.45}_{-0.29} \pm 0.11 ~,~~ 1.1 \le q^2 \le 6.0 ~{\rm GeV}^2 ~, \\
0.90^{+0.27}_{-0.21} \pm 0.10 ~,~~ 0.1 \le q^2 \le 8.0 ~{\rm GeV}^2 ~, \\
1.18^{+0.52}_{-0.32} \pm 0.10 ~,~~ 15.0 \le q^2 \le 19.0 ~{\rm GeV}^2 ~, \\
0.94^{+0.17}_{-0.14} \pm 0.08 ~,~~ 0.045 \le q^2 ~.
\end{cases} \nn
\eeq
Although the central values are closer to the SM predictions, the
errors are considerably larger than in the LHCb measurement.

\item[$\bullet$] On average, older measurements of the branching ratio of $B_s
  \to \mu^+ \mu^-$ were in agreement with the prediction of the SM
  \cite{Chatrchyan:2013bka,CMS:2014xfa,Aaij:2017vad}.  However, a
  new measurement by ATLAS disagrees with SM by $2.4\sigma$
  \cite{Aaboud:2018mst}. Combining all results leads to tension of
  $\sim 2\sigma$ with the SM.

\end{itemize}

There are therefore quite a few measurements of observables that are
in disagreement with the predictions of the SM. All of these involve
the decay $\bsmumu$, which suggests trying to explain the data by
allowing NP to contribute to this decay. The model-independent
starting point is the effective Hamiltonian
\beq
H_{\rm eff} = - \frac{\alpha G_F}{\s \pi} V_{tb} V_{ts}^*
      \sum_{a = 9,10} ( C_a O_a + C'_a O'_a ) ~,
\eeq
where $O_{9(10)} = [ {\bar s} \gamma_\mu P_L b ] [ {\bar\mu}
  \gamma^\mu (\gamma_5) \mu ]$, and the primed operators have $L \to
R$. The Wilson coefficients include both SM and NP contributions: $C_X
= C_{X,\SM} + C_{X,\NP}$.

Performing a combined fit to all the data, in the simplest scenarios
it is found that the data can be explained if\footnote{These numbers
  are taken from Ref.~\cite{Datta:2019zca}.  Other analyses
  \cite{Alguero:2019ptt,Ciuchini:2019usw,Aebischer:2019mlg,Kowalska:2019ley,Arbey:2019duh}
  obtain similar results.}
\bea
  {\rm (i)} ~~~~~ C^{\mu\mu}_{9,\NP} &=& -1.10 \pm 0.16 ~, \nn\\
{\rm (ii)} ~~~~~ C_{9,\NP}^{\mu\mu} &=& -C_{10,\NP}^{\mu\mu} = -0.53 \pm 0.08 ~,
\label{scenarios}
\eea
with a pull of close to $6\sigma$(!). (I note in passing that scenario
(ii) involves purely left-handed NP.)

\subsection{$\bctaunu$}

\begin{figure}[t]
\begin{center}
\includegraphics[width=0.4\hsize]{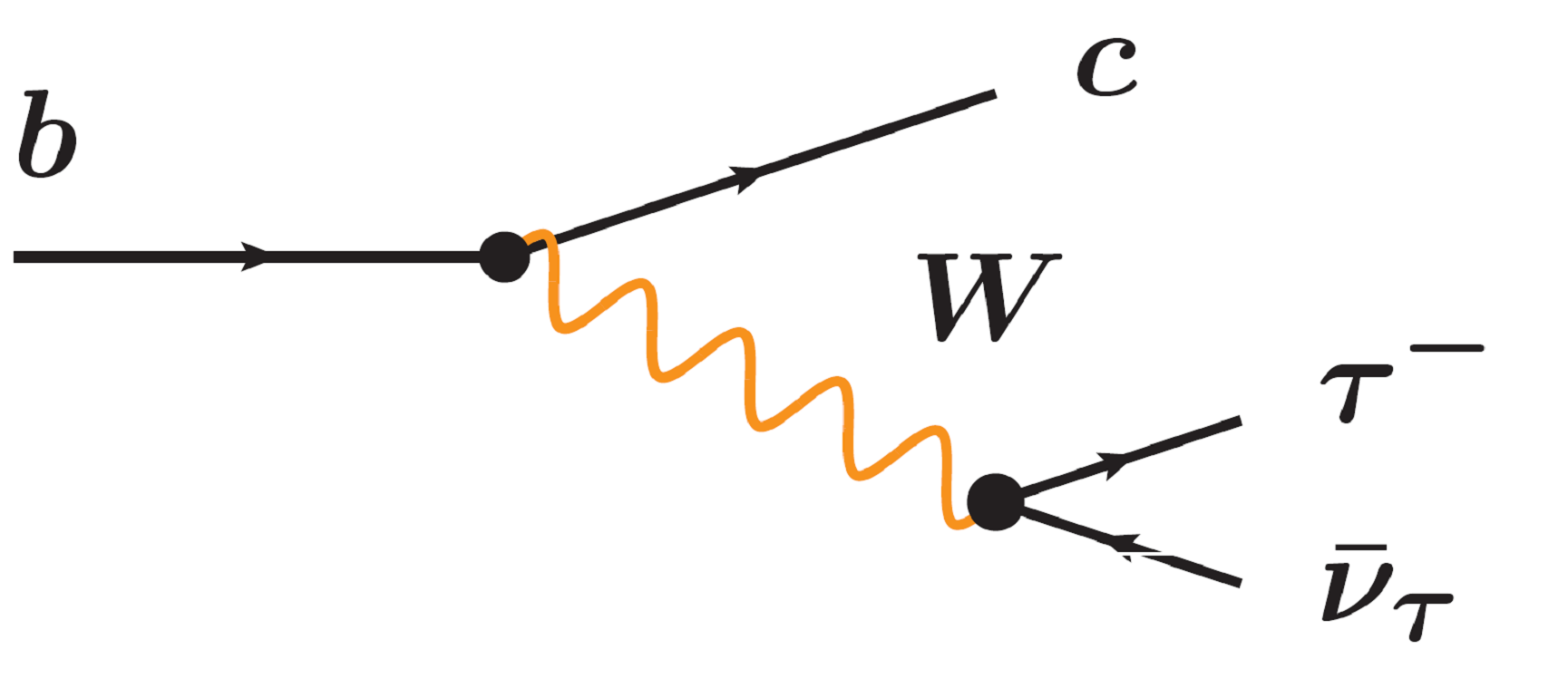}
\caption{SM diagram for $\bctaunu$.}
\label{bctaunufig}
\end{center}
\end{figure}

There is another set of observables whose measurements also exhibit
discrepancies with the SM. They involve the decay $\bctaunu$. This is
a $\Delta Q_{em} = 1$ process, and proceeds in the SM via tree-level
$W$ exchange, see Fig.~\ref{bctaunufig}. The amplitude is given by
\beq
A \sim \frac{g_2^2}{M_W^2} \, V_{cb} ~,
\eeq
where $|V_{cb}| \simeq 0.04$.

Before the Rencontres de Moriond, 2019, BaBar, Belle and LHCb measured
the quantities
\beq
R_{D^{(*)}} \equiv \frac
{\cB(\bar{B} \to D^{(*)} \tau^{-} {\bar\nu}_\tau)}{\cB(\bar{B} \to D^{(*)} \ell^{-} {\bar\nu}_\ell)} ~~,~~
  (\ell = e,\mu) ~. 
\eeq
Their measurements exhibited discrepancies with the predictions of the
SM. Combining $R_D$ and $R_{D^*}$, the deviation was $\sim
3.8\sigma$ \cite{HFLAV}. At Moriond, 2019, Belle announced new results
\cite{Abdesselam:2019dgh}:
\bea
R_{D^*}^{\tau/\ell}/(R_{D^*}^{\tau/\ell})_\SM &=& 1.10 \pm 0.09 ~, \nn\\
R_{D}^{\tau/\ell}/(R_{D}^{\tau/\ell})_\SM &=& 1.03 \pm 0.13 ~.
\eea
These results are in better agreement with the SM, so that the
deviation from the SM in $R_D$ and $R_{D^*}$ (combined) has been
reduced from $\sim 3.8\sigma$ to $3.1\sigma$ \cite{HFLAV}.

LHCb has also measured 
\beq
R_{J/\psi} \equiv \frac{\cB(B_c^+ \to J/\psi\tau^+\nu_\tau)}{\cB(B_c^+ \to J/\psi\mu^+\nu_\mu)} ~,
\eeq
finding \cite{Aaij:2017tyk}
\beq
\frac{R_{J/\psi}}{(R_{J/\psi})_\SM} = 2.51 \pm 0.97 ~.
\eeq
Here the discrepancy with the SM is $1.7\sigma$ \cite{Watanabe:2017mip}.

The discrepancies in $R_{D}$, $R_{D^*}$ and $R_{J/\psi}$ are hints of
$\tau$-$\mu$ and $\tau$-$e$ universality violation in $b \to c \ell^-
{\bar\nu}$, and suggest the presence of NP in $\bctaunu$ decays.

\section{Models of New Physics}

For the $\bsmumu$ anomalies, there are two classes of NP models that
contribute to the decay at tree level, and can explain the data. 

\begin{figure}[t]
\begin{center}
\includegraphics[width=0.4\hsize]{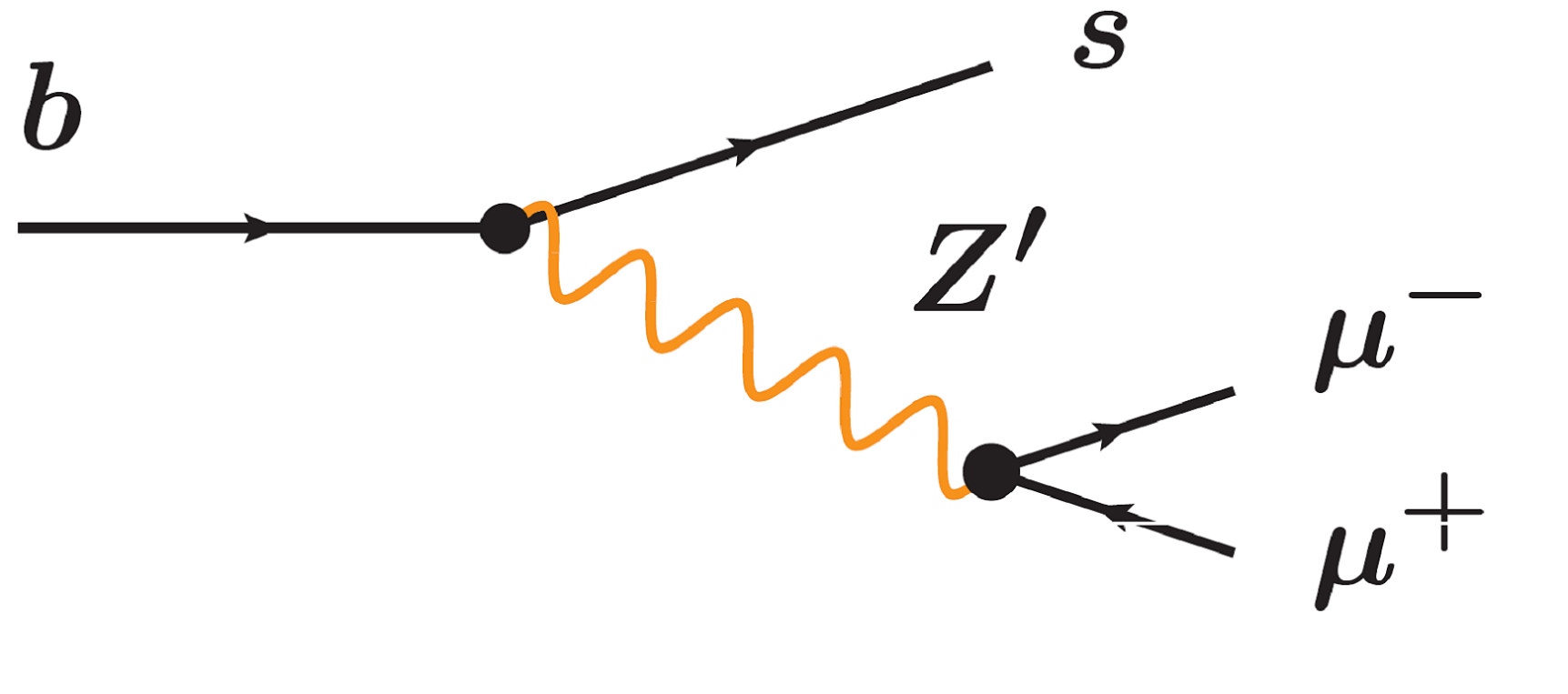}
\hskip1truecm
\includegraphics[width=0.4\hsize]{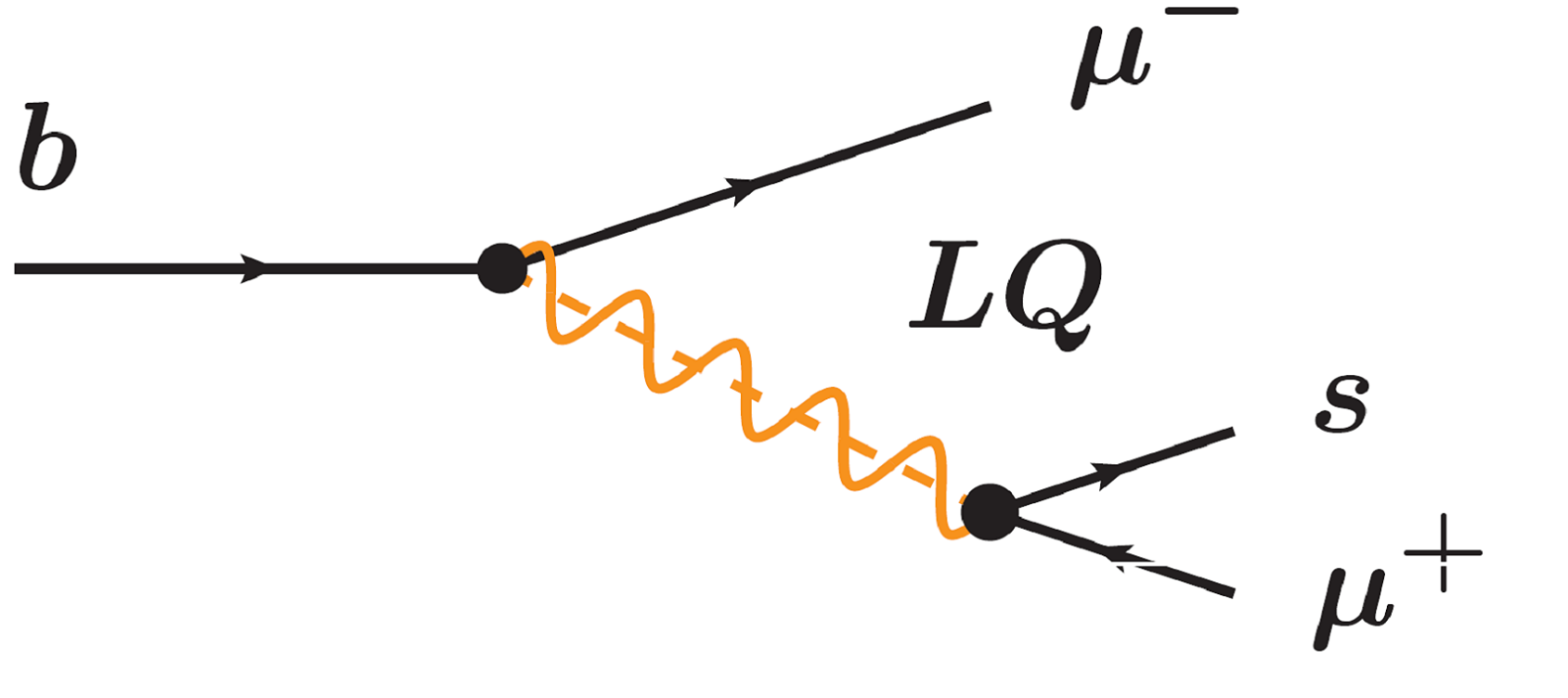}
\caption{$Z'$ (left) and LQ (right) contributions to $\bsmumu$.}
\label{bsmumuNP}
\end{center}
\end{figure}

The first class involves a new $Z'$ boson (see Fig.~\ref{bsmumuNP}).
The $Z'$ must have a FCNC coupling to ${\bar s} b$ and must couple to
$\mu^+ \mu^-$. The model can follow scenarios (i) or (ii)
[Eq.~(\ref{scenarios})]. A great many $Z'$ models have been proposed
(far too many to list here). Some combine explanations of the $B$
anomalies with other weaknesses of the SM, such as dark matter,
$(g-2)_\mu$ and neutrino masses.

\begin{figure}[t]
\begin{center}
\includegraphics[width=0.4\textwidth]{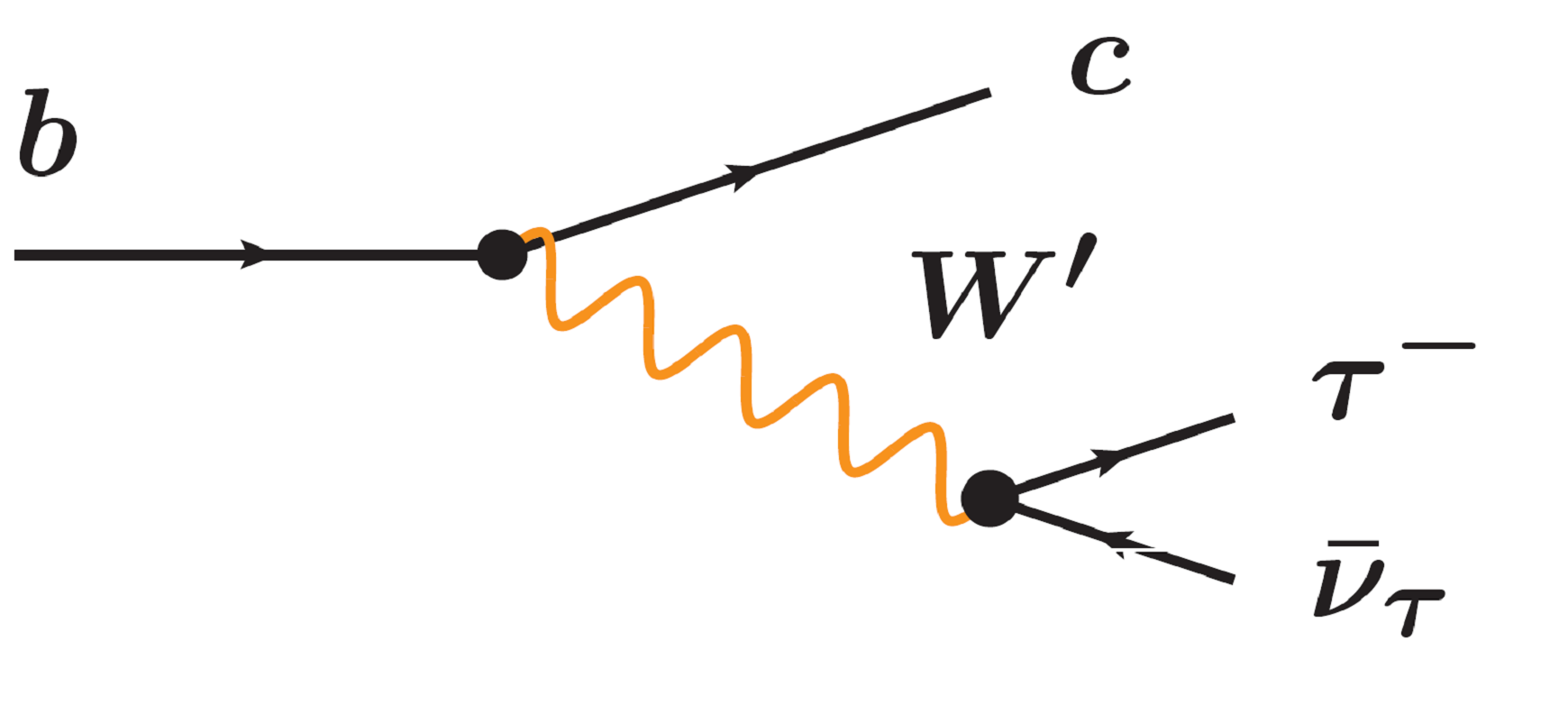}
\hskip1truecm
\includegraphics[width=0.4\textwidth]{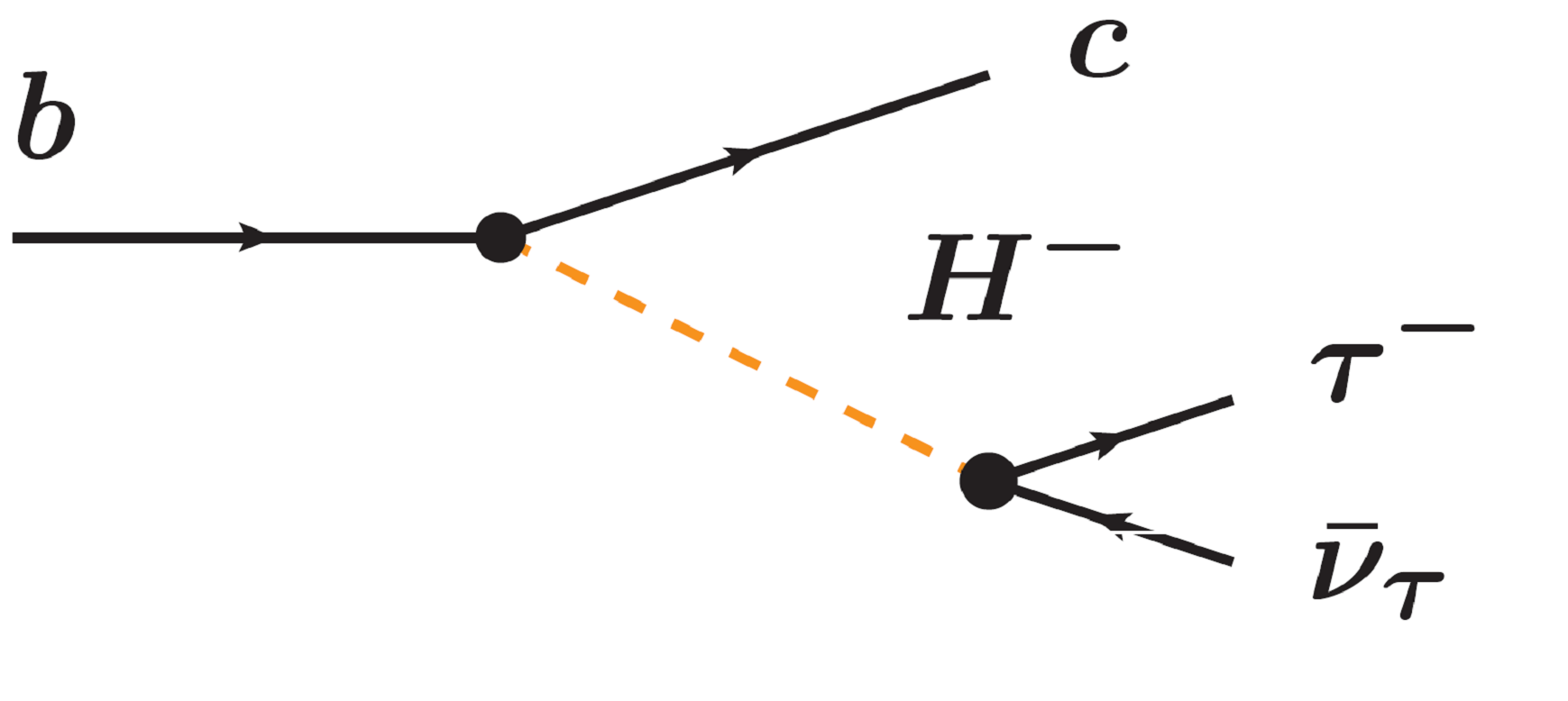} \\
\includegraphics[width=0.4\hsize]{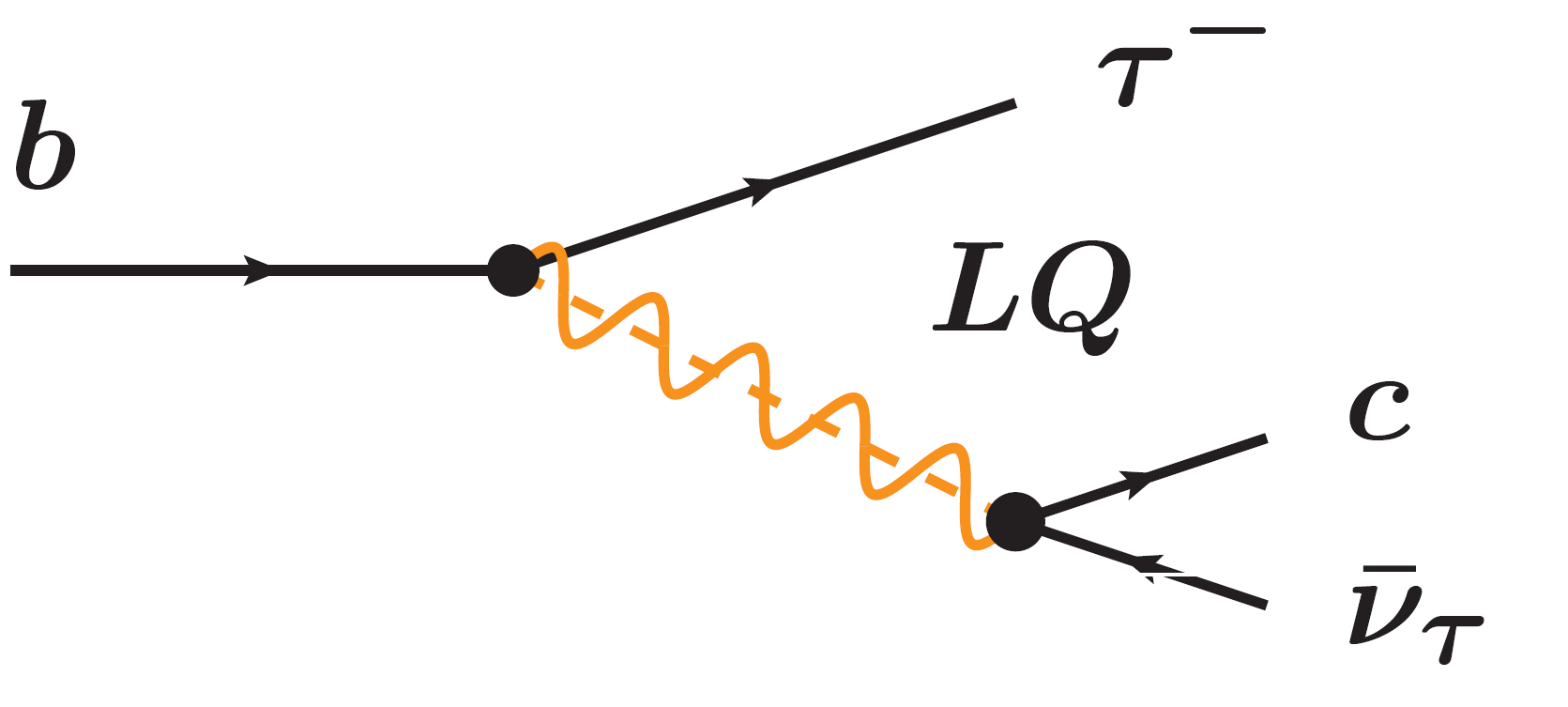}
\caption{$W'$ (top left), charged-Higgs (top right) and LQ (bottom)
  contributions to $\bctaunu$.}
\label{bctaunuNP}
\end{center}
\end{figure}

The second class of NP models involves leptoquark (LQ) exchange (see
Fig.~\ref{bsmumuNP}).  There are several different types of LQ that
can explain the $\bsmumu$ data. All fit within scenario (ii) of
Eq.~(\ref{scenarios} (purely LH NP) \cite{Alok:2017jgr}.

Turning to $\bctaunu$, there are three types of NP whose contributions
to this decay could explain the data: (1) a new $W'$ boson, (2) a
charged Higgs boson, and (3) a leptoquark (see
Fig.~\ref{bctaunuNP}). However, the $H^-$ is disfavoured by
(theoretical) constraints from $B_c^- \to \tau^- {\bar\nu}_\tau$
\cite{Alonso:2016oyd}, leaving the $W'$ or (several different types
of) LQ as NP explanations. Here, the NP couplings can be left-handed
(LH) and/or right-handed (RH).

\subsection{Distinguishing NP Explanations}

As we have seen, there are several NP explanations for the anomalies
in $\bsmumu$. But this raises the question: how can we distinguish
among them?  One way is to look at CP violation in $B \to K^* \mu^+
\mu^-$ \cite{Alok:2017jgr}. Now, CP violation is generated by the
interference of (at least) two amplitudes with different weak phases.
In the presence of NP, this can arise due to SM-NP interference. Here
the signal is not direct CP violation, but rather CP asymmetries in
the angular distribution. The key point here is that SM-$Z'$ and SM-LQ
interferences are different, leading to different CP-violating
effects. Thus, by measuring CP violation in $B \to K^* \mu^+ \mu^-$,
one can differentiate the NP models.

The situation is similar for the NP explanations of the $\bctaunu$
anomalies.  By looking at CP violation in $B \to D^* \tau^- {\bar\nu}$
(and also in $B \to D^* \mu^- {\bar\nu}$), one can distinguish NP
models \cite{Bhattacharya:2019olg}. Once again, the signal involves CP
asymmetries in the angular distribution. The measurement of CP
violation in these decays allows us to differentiate the $W'$ and LQ
models. It also provides information about the LH/RH NP couplings.

\subsection{Simultaneous explanations of $\bsmumu$ and $\bctaunu$}

Now, $(c,s)_L$ is a doublet of $SU(2)_L$. This suggests that, if the
NP coupling is purely LH, $b \to s$ and $b \to c$ transitions are
related. It should therefore be possible to find NP that can
simultaneously explain both the $\bsmumu$ and $\bctaunu$ anomalies
\cite{Bhattacharya:2014wla}.

There are two classes of models that, in principle, can do this:
\begin{enumerate}

\item[$\bullet$] A new triplet of vector bosons $(W^{\prime \pm},Z^{\prime 0})$.
  The $W'$ and $Z'$ contribute respectively to $\bctaunu$ and
  $\bsmumu$.

\item[$\bullet$] A LQ of charge $Q_{em} = \frac23$. It couples to ${\bar b}
  \mu^+$ and ${\bar s} \mu^+$ (for $\bsmumu$) and to ${\bar b} \tau^+$
  and $c {\bar\nu}_\tau$ (for $\bctaunu$).

\end{enumerate}

It is found \cite{Bhattacharya:2016mcc,Buttazzo:2017ixm,Kumar:2018kmr,Cornella:2019hct} that, when all constraints are
taken into account, including those from direct searches at the LHC,
the $(W^{\prime \pm},Z^{\prime 0})$ model is excluded. But the LQ
model is viable!

\section{Summary}

The SM is certainly correct, but it is not complete: there must be
physics beyond the SM. Recently, there have been several measurements
of observables that are in disagreement with the predictions of the
SM:
\begin{itemize}

\item[$\bullet$] $\bsmumu$: These include many observables involving
  this decay.  Some are clean, while others have important theoretical
  uncertainties.  Global fits allowing for NP in $\bsmumu$ find
  improvements over the SM at the level of close to $6\sigma$. NP
  Models with an extra $Z'$ or with different types of LQs have been
  proposed as explanations.

\item[$\bullet$] $\bctaunu$: here there are several clean observables, with a net
  deviation from the SM of $\sim 3\sigma$. These can be explained in
  models with an extra $W'$ or with different types of LQs.

\end{itemize}

It is of course possible that these discrepancies with the SM are all
statistical fluctuations, and will go away with more data.  This said,
their combined statistical significance is sizeable ($\gsim 4\sigma$),
so they will not disappear soon. Hopefully, we are indeed seeing the
first experimental signals of NP.

\bigskip
\noindent
{\bf Acknowledgments}: This work was financially supported in part by
NSERC of Canada.

%

\end{document}